# $R_K/100$ and $R_K/200$ Quantum Hall Array Resistance Standards

W. Poirier, A. Bounouh, K. Hayashi, H. Fhima, F. Piquemal, G. Genevès
*Bureau National de Métrologie-Laboratoire National d'Essais,*
*33 avenue du Général Leclerc F-92260 Fontenay-Aux-Roses, France*

J. P. André
*Laboratoire d'Electronique de Philips (LEP), F-94451 Limeil-Brévannes, France*

## Abstract

It is theoretically possible to combine several Hall bars in arrays to define new quantum standards with perfectly quantized resistance values. We have thus, for the first time, developed and fabricated novel Quantum Hall Array Resistance Standards (QHARS) made of a large number $N$ ($N$=100, 50) of Hall bars placed in parallel using a triple connections technique. The Hall resistance of these quantum standards is found to be very well quantized. On the $i$=2 Hall plateau, the resistance of specific good arrays stays equal to $R_K/2N$ within 5 parts in $10^9$ for supplying currents up to 2 mA at a temperature of 1.3 K. The mean longitudinal resistance of the Hall bars which constitute the arrays has been determined through the analysis of the array equivalent electrical circuit. This measurement shows that the carrier transport in the Hall bars is dissipationless. This work therefore demonstrates the efficiency of the multiple connections technique and consequently that QHARS are likely to extend the QHE metrological applications.

## I. Introduction

After the discovery of the Integer Quantum Hall Effect (IQHE) in 1980 by K. von Klitzing[1], a lot of experimental work showed that this effect is sample-independent. The Hall resistance of a two-dimensional electron gas in a perpendicular magnetic field is quantized in integer fractions of the universal constant $R_K \equiv h/e^2$ independently of the electronic density or mobility values and disorder nature. Laughlin[2] and then Halperin[3] brought a general explanation to these singular properties based on gauge invariance and the existence of a mobility gap. The universality and the reproducibility of this effect led the Comité International des Poids et Mesures (CIPM) to recommend the IQHE in 1990 for maintaining the unit of resistance in the National Metrology Institutes (NMIs)[4]. The $R_K$ resistance value can indeed be reproduced within some parts in $10^{10}$. However, a relative uncertainty of one part in $10^7$ is assigned to its 25812,807 Ω value. Although SI direct measurements[6,7] of $R_K$ via calculable capacitors have been obtained with uncertainties as low as 2.4 parts[8,9] in $10^8$, the CIPM established the value of $R_K$ with a larger uncertainty because, assuming that $R_K$ is presumably equal to $h/e^2$, it took into account indirect determinations of $h/e^2$ and calculations of the fine structure constant. That still remains a big metrological stake to prove that the relation $R_K=h/e^2$ is true with a better accuracy. In the meantime, the conventional value $R_{K-90}$=25812.807 Ω is used to insure the consistency of resistance measurements all over the world[4].

Although maintaining $R_K$ within one part in $10^9$ is achievable by respecting some technical guidelines[10], the use of the IQHE has some limitations. Only $R_K/4$ and $R_K/2$ resistance values can be reasonably operated in metrological applications. Plateaus corresponding to odd or high value quantum numbers are not well quantized. Moreover, the maximum current that can be supplied in classical Hall bars is limited by strong dissipation in the contact corners[11]. Special resistance bridges whose use is restricted to NMIs are consequently required to calibrate material standards with IQHE. Fortunately, some experiments have demonstrated the possibility to obtain new quantum resistance values. Syphers, Fang and Stiles[12] firstly showed that any rational fraction of the two-terminal resistance could be attained by interconnecting several samples with the appropriate connections. They also proved that a similar result could be obtained with multiple connections on a single sample[13]. Then, F. Delahaye[14] proposed a new multiple connections technique with redundant links between multi-terminal QHE devices connected in series or in parallel. This technique allows one to cancel the contact resistance effect and consequently to define the four-terminal resistance of the equivalent quantum resistors. It is based on the Ricketts and Kemeny[15] electrical representation of Hall bars in the quantized state. Expressing the typical resistance of a contact between two Hall bars relative to the value of $R_H$ by $\varepsilon$ ($\varepsilon \ll 1$), the relative error contribution to the Hall resistance is limited to $\varepsilon^n$, where n is the number of links. This technique uses two fundamental properties of the IQHE: the two-terminal resistance between any pair of probes and the four-terminal longitudinal resistance are respectively and ideally equal to $R_H$ and zero. In case of Hall chemical potential differences much smaller than the cyclotron energy ($eV_H \ll \hbar\omega_c$) the edge-states theory[3,16] has been particularly able to enlighten these two properties. It describes dissipationless and one-dimensional currents flowing along the sample edges between reservoirs (contacts) where dissipation randomizes the wave-function phase. The equipotentiality is related to the absence of back-scattering due to the spatial separation of edge states with opposite momentums. In case of edge states with perfect transmission, the two-terminal resistance is consequently equal to an integer fraction of $h/e^2$ and the four-terminal longitudinal resistance is zero. The other contribution of the Büttiker's paper[16] is the explanation of the existence of the IQHE even in large open samples like those used in metrological applications where inelastic scattering takes place. Phase rigidity of the wave function is indeed only needed over cyclotron radius length. For higher Hall voltages, a lot of experiments[17-20] show that a large fraction of the current carried by bulk states circulates inside the sample contrary to what is predicted by the edge-states model. A complete theory of the IQHE taking into account high currents is always in expectation. In any case, there is experimental evidence that the two properties described before remain valid up to a supplying critical current corresponding to excitations of magnitude much higher than the cyclotron energy.

The multiple connections technique proposed for metrological applications therefore allows us to build artificial macroscopic Hall samples while keeping the fundamental properties of the IQHE. The experimental proof of the efficiency of the multiple connections technique for metrological applications was firstly obtained for two Hall bars placed in parallel[14,21], then for Hall bars connected in auto-series[22]. We have successfully realized the first standard made of ten Hall bars connected in series by triple links[22]. The next step consists in checking the efficiency of the multiple connections technique for standards based on a large number of Hall bars. The availability of Quantum Hall Array Resistance Standards (QHARS) with resistance values from 100 Ω to 1 MΩ is indeed likely to open new prospects.

NMIs will be able to reduce their measurement uncertainties by shortening the calibration chain of their material standards. For example, a $R_K/200$ QHARS could allow a direct calibration of 1 Ω or 10 kΩ resistance standard without using a 100 Ω transfer material resistor. A real accuracy improvement of the high resistance standard calibration is also expected by starting the measurement chain from $50R_K$ (~ 1.29 MΩ) rather than 10 kΩ.

The recent signature of the Mutual Recognition Arrangement[23] (MRA) of national measurement standards and of calibration and measurement certificates issued by national institutes has reinforced the necessity to check the consistency of the resistance unit realizations at different NMIs. International comparisons are thus playing a growing role. The availability of good comparison standards is consequently a big stake. But the comparison of QHE resistance bridges at a level of some parts in $10^9$ is very difficult with material standards: resistors suffer random resistance changes of some parts in $10^8$ during transportation that are not always completely cancelled even after a long recovery time. The use of specially designed pressure and temperature stabilized resistor is a way to avoid these problems[24]. However, this kind of standard is rare. Direct comparison of QHE set-ups obviously constitutes a very accurate[21] but nevertheless cumbersome calibration method. QHARS used as traveling standards might solve a lot of these problems. Because they are handy and not dependent on transportation conditions, quick and very accurate international comparisons might be possible.

Finally, the development of QHARS might generalize the maintaining of the unit of resistance with QHE. Until now, the use of QHE at the best level is limited to important NMIs because of the need of quite complex and expensive equipment based on the cryogenic current comparator[25-27] (CCC). Henceforth, samples characterized by a low density of electrons and a high mobility may offer a first simplification of the measurement system since they work with lower magnetic field and higher temperature[28,29]. However, the CCC stays essential if using classical Hall bars supplied by currents limited to some tens of µA. But with QHARS specially designed to work with the high current required by conventional resistance bridges, any NMIs, and even calibration centers, might be more easily equipped with QHE standards.

In this paper, we present the first realization of QHARS based on the connection in parallel of a large number of Hall bars. Quantum standards of expected nominal resistance values equal to $R_K/200$, $R_K/100$, $R_K/50$, $50R_K$ ($i=2$ plateau) have been fabricated in collaboration with the LEP. Only studies related to $R_K/200$ (QHARS129) and $R_K/100$ (QHARS258) nominal resistance arrays processed on PL174 AlGaAs/GaAs heterostructures[22] are described here. Metrological characteristics and Hall resistance measurements by means of direct and indirect comparisons with $R_K$ are given: we analyse the contact resistance, current and temperature effects on the quantization of the equivalent Hall resistance. Our work shows that the Hall resistance of QHARS agrees with their expected nominal value within 5 parts in $10^9$. This high level of quantization is directly related to the efficiency of the triple connection which is explained through the equivalent electrical circuit analysis of the arrays. We finally propose new ideas to improve the design of QHARS.

## II. Arrays and Experimental setup

Measurements presented here were performed on a two-dimensional electron gas obtained from AlGaAs/GaAs heterostructures (PL174) grown on 3-in wafers by metal-organic vapor-phase epitaxy (MOVPE) process. Starting from the substrate, a 600 nm thick un-doped GaAs buffer layer was deposited followed by a 14.5 nm thick un-doped $Al_{0.28}Ga_{1-0.28}As$ spacer layer. Then a 40 nm thick $10^{18}$ cm$^{-3}$ Si-doped $Al_{0.28}Ga_{1-0.28}As$ layer was realized. Finally n-type 12 nm GaAs cap layer covers the heterostructures to improve ohmic contact.

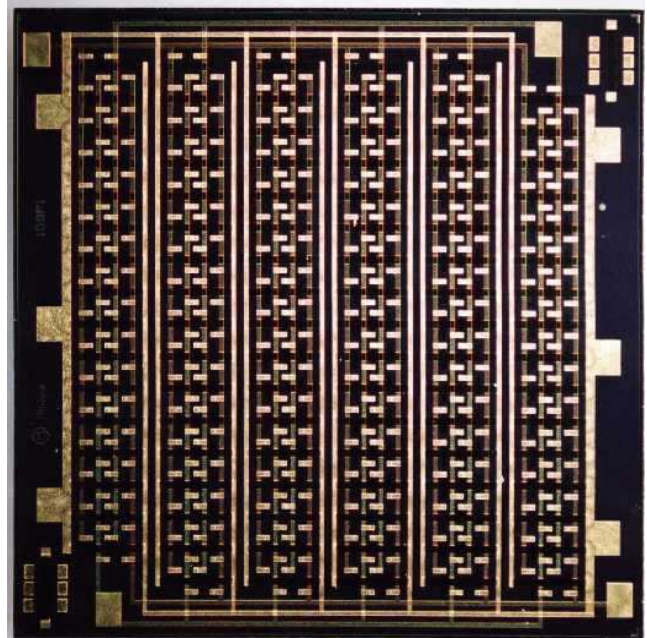

Fig. 1 : picture of a QHARS129 sample.

To realize such devices, 100 or 50 Hall bars are placed in parallel by triple connections[14]. Current terminals and four potential terminals of the different Hall bars are respectively connected between them by gold paths. The six electrical circuits built in this way are isolated from each other. Fig. 1 shows a picture of a QHARS129 sample. Schematic views of the connections are presented in Fig. 2a and Fig. 2b. The technological process has more steps than for the fabrication of 10-QHE devices series array resistance standards[22]. Firstly, a 300 nm thick mesa delimiting the active area is realized. AuGeNi ohmic contacts are then evaporated and annealed at 450°C (this stage being the most crucial for the metrological applications). The first level of insulation is ensured by a 150 nm thick $Si_3N_4$ layer, which allows us to reduce the mechanical strains exerted on the 2DEG. Two levels of gold paths separated by a 300 nm thick $S_iO_2$ insulating layer are then deposited. Inter-level connections are realized by etching processes. Fig. 2c shows a schematic cross section of the layers. The overall dimensions of each chip are restricted to 7.5×7.5 mm$^2$ in respect to the size of EUROMET TO-8 holders. Therefore, the length L and the width W of Hall bars of QHARS129 and QHARS258 samples are reduced to 950 and 1100 µm and to 100 and 200 µm respectively. The length l between voltage probes is 200 µm (see Fig. 2b). Let us note that the reduction of the individual bar dimensions unfortunately reduces the critical current and working current values.

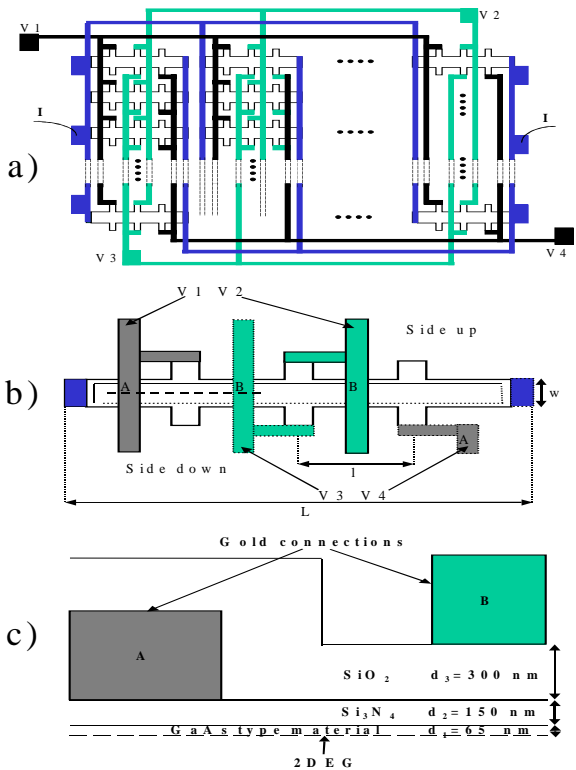

Fig. 2: schematic views of the connections: a) top view of the sample; b) enlarged top view of one particular Hall bar; c) cross section showing the different layers.

Our experiment includes a cryostat fitted with a pumped $^4$He insert to reduce the temperature inside the sample chamber below 1.3 K, and contains a super-conducting magnet system which produces a maximum magnetic flux density of 14 T at 4.2 K. Our resistance ratio bridge represented in Fig. 3 operates at DC using a CCC coupled to a RF SQUID by a super-conducting flux transformer. The two four-terminal standard resistors $R_P$ and $R_S$ are connected respectively to the primary ($N_P$ turns) and the secondary ($N_S$ turns) windings of the CCC chosen so that the $N_P/N_S$ ratio is near $R_P/R_S$. These resistors are supplied with a double current source ($I_P$ and $I_S$ respectively). The fine regulation of the current ratio is ensured by the RF SQUID working in the external feedback mode to realize zero flux detection[30,31]. A resistive divider is adjusted to supply an auxiliary winding of the CCC with a $\varepsilon \times I_S$ small fraction of the $I_S$ current in order to obtain zero voltage at the null detector simultaneously. The balance of the bridge consequently yields the following equations :

$$R_P \times I_P = R_S \times I_S \qquad (1)$$

$$N_P \times I_P = N_S \times I_S \times (1 + \varepsilon \times N_A / N_S) \qquad (2)$$

which give the resistance ratio:

$$\frac{R_S}{R_P} = \frac{N_S}{N_P} \times \left(1 + \varepsilon \times \frac{N_A}{N_S}\right) \qquad (3)$$

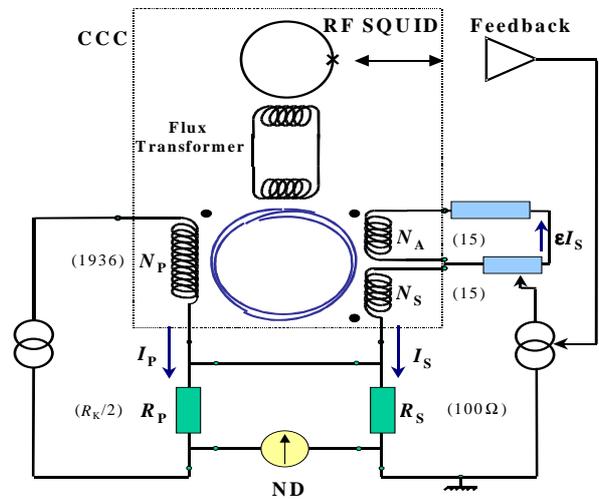

Fig. 3: schema of the cryogenic resistance bridge.

Usually this resistance ratio bridge is able to give a total relative uncertainty on the 100 and 129 resistance ratios less than 2×10$^{-9}$. The Hall resistance of QHARS samples can be directly compared to that of a single Hall bar (PL175[32] or PL174 samples) ($R_K/2$) measured on the $i$=2 plateau and placed in the same cryostat. Indirect comparisons can also be performed via 1 Ω or 2 Ω material resistors calibrated independently in terms of $R_K/2$ through an ultra stable 100 Ω resistor. Fig. 4 gives a schematic of the measurement chains. Finally, all the measurements described here have been obtained in a 1.3 K to 4.2 K temperature range.

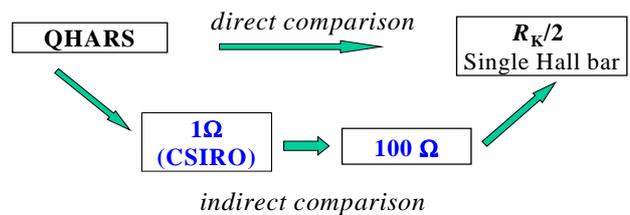

Fig. 4: schema of the measurement chains.

## III. Equivalent circuit

An equivalent circuit of the array reproducing the specific properties of QHE is needed to understand its electrical properties. Fig. 5 shows the circuit used to describe the sample. It is based on a six-terminal 'diamond' Hall bar model[15] including contacts ($\varepsilon \ll 1$) and longitudinal ($\beta \ll \varepsilon \ll 1$) resistance contributions expressed in units of half the Hall resistance of a single bar $r_H/2$. Each line corresponds to one Hall bar. $N$ Hall bars numbered by the subscript q are connected in parallel. A star connection type is assumed between terminals: the resistance of the links being introduced in the $\varepsilon$ quantity. The direction of the magnetic field indicated in Fig. 5 corresponds to the positive direction ($B+$) in the discussion. Of course, this model reproduces the ideal and macroscopic QHE properties only. In case of zero $\varepsilon$ and $\beta$ values, it notably accounts for the following properties: for any single Hall bar, the two terminal resistance is equal to $r_H$, while the four terminal longitudinal and transverse resistance values are equal to zero and $r_H$ respectively. On the other hand, it doesn't describe the fine microscopic behavior of the electron gas: for example, the effect of the contact resistance on edge channel equilibrium is not taken into account. It has been shown that the non-equilibrium distribution, which may be responsible for Hall resistance deviations, can travel over macroscopic distances (> 100 µm)[33]. However, these results only concern transport in small systems with $eV_H \ll \hbar\omega_c$. For the Hall bars described in this paper, $\hbar\omega_c$ is about 16 meV at the center of the $i=2$ plateau. And, it has been shown that for Hall bars used for the metrological application, supplied with currents above 10 µA ($eV_H > 130$ meV on the $i=2$ plateau) at a temperature above 0.3 K, relative deviations of the Hall resistance less than one part in $10^9$ are assumed if the voltage contact resistance values are less than 100 Ω[34,35]. Since the deviations exponentially decay with the temperature, and because contact resistance values much lower than 100 Ω are usually found with the LEP technological process, smaller deviations are therefore expected in our case. On the other hand, the in-homogeneity of the contact or longitudinal resistance values implies that currents circulate in the different connections. The $i_q, i'_q, i''_q, j_q, j'_q, j''_q$ currents are indicated in Fig. 5. The following notations are used:
$R_H^C = |V2-V3|/I$, $R_H^D = |V1-V4|/I$, $R_{xx}^a = |V1-V2|/I$ (or $|V3-V4|/I$).

The case $\beta_q=0$ has been particularly described for two Hall bars by F. Delahaye[14]. We have introduced the longitudinal resistance not only to determine the quantity $R_{xx}^a$ but also to evaluate additional contributions to the deviation of the Hall voltages from their nominal values. The in-homogeneity of the electronic and magnetic flux densities over the large size of the arrays may indeed be responsible for larger $\beta_q$ values than in the case of classical Hall bars. Let us firstly describe the qualitative behavior of the system by considering the case of only two samples with opened voltage links. The inequality of the current contact resistance values $\varepsilon_{11a}$ and $\varepsilon_{21a}$ create a voltage between the points C and D. As the resistance seen from these points is approximately equal to $2r_H$, closing this loop generates circulation of a current $i'_1 (= -i'_2)$ proportional to the difference ($\varepsilon_{21a}-\varepsilon_{11a}$). The additional voltage drop due to $\varepsilon_{11b}$ and $\varepsilon_{21b}$ terms is therefore qualitatively proportional to $\varepsilon^2$. Finally, this voltage plus the potential due to the inequality of the longitudinal resistance values $\beta_1$ and $\beta_2$ generate a current $i''_1 (= -i''_2)$ in the closed connection EF of the qualitative form $\varepsilon^2+\beta$. The additional voltage drop along the third connection which adds to the Hall voltage is consequently of the form $\varepsilon^3+\beta\varepsilon$. Starting now with the resistance differences on the right side, the same reasoning allows to determine the currents $j'_1, j''_1$. The general method of calculation for any number N of Hall bars connected by triple connections is detailed in the Appendix. The equations are solved by neglecting terms in $\varepsilon$ and $\beta$ of orders higher than 3. The expression of any quantity depends on all the $\varepsilon_q$ and $\beta_q$ parameters. Table I gathers all the qualitative results.

| Quantities | $\varepsilon=0$ $\beta=0$ | $B+$ | $B-$ |
|---|---|---|---|
| $R_H^C$ | $\dfrac{r_H}{N}$ | $\dfrac{r_H}{N}\left[1+O(\beta\varepsilon,\varepsilon^3)\right]$ | $\dfrac{r_H}{N}\left[1+O(\beta\varepsilon,\varepsilon^2)\right]$ |
| $R_H^D$ | $\dfrac{r_H}{N}$ | $\dfrac{r_H}{N}\left[1+\overline{\beta}+O(\beta\varepsilon,\varepsilon^2)\right]$ | $\dfrac{r_H}{N}\left[1-\overline{\beta}+O(\beta\varepsilon,\varepsilon^3)\right]$ |
| $R_{xx}^a$ | 0 | $\dfrac{r_H}{N}\left[\dfrac{\overline{\beta}}{2}+O(\beta\varepsilon,\varepsilon^2)\right]$ | $\dfrac{r_H}{N}\left[\dfrac{\overline{\beta}}{2}+O(\beta\varepsilon,\varepsilon^2)\right]$ |

Table I: sum up of qualitative calculated deviations.

Results for positive and negative magnetic field are different because the geometry of the connections is not invariant by magnetic field reversal. Briefly, $R_H^C(B+)$ and $R_H^D(B-)$ are modified by only $\varepsilon^3$ terms while $\varepsilon^2$ corrections add to $R_H^C(B-)$, $R_H^D(B+)$ and $R_{xx}^a$. In addition, our calculations show that the sample is characterized by a non zero longitudinal resistance $R_{xx}$ whose first order term in $\beta$ is given by (see equation 27):

$$R_{xx} = \frac{r_H}{2N}\overline{\beta} = \frac{\overline{r_{xx}}}{N}, \quad (4)$$

$\overline{r_{xx}}$ being the mean value of the Hall bar longitudinal resistance values. This contribution appears in $R_H^D$ and $R_{xx}^a$ quantities. But the best way to measure $R_{xx}$ consists in comparing $R_H^D(B-)$ and $R_H^C(B+)$. Let us consider $R_{xx}^b$ defined by:

$$R_{xx}^b = [R_H^C(B+) - R_H^D(B-)]/2. \quad (5)$$

Contrary to $R_{xx}^a$, $\varepsilon^2$ terms don't contribute to that quantity, which can be written in the following way:

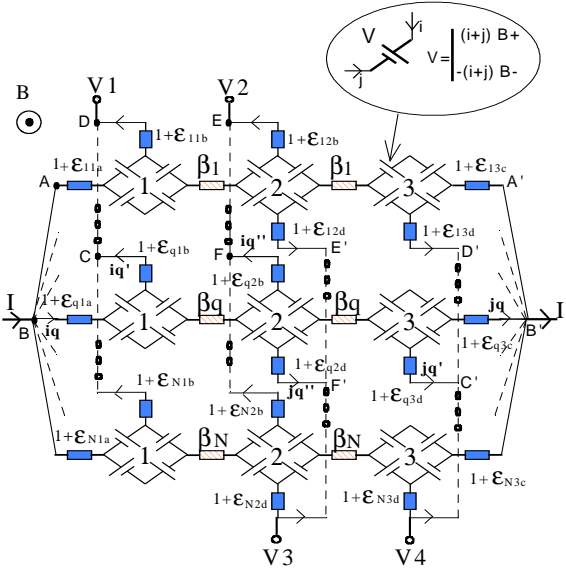

Fig. 5: array equivalent electrical circuit. Resistance and voltage values are expressed in terms of $r_H/2$.

$$R_{xx}^{b} = R_{xx} + O(\beta\varepsilon, \varepsilon^3, \varepsilon\beta^2, \beta\varepsilon^2, \beta^2). \quad (6)$$

Finally, let us note that the introduction of the longitudinal resistance shows that a $\beta\varepsilon$ contribution adds to any quantity. These terms are usually smaller than $\varepsilon^2$ or $\varepsilon^3$ terms. But their contributions however prevail when the number of connections increase. They therefore limit the efficiency of the multiple connections technique. In the following $\beta$, $R_{xx}^{a}$, $R_{xx}^{b}$ and $V_{xx}^{a}$ refer to the measured quantities multiplied by the ratio W/l (see Fig. 2b).

## IV. Results

Practical realizations of quantized resistance values with samples made of 50 or 100 Hall bars placed in parallel on area as large as 1 cm$^2$ assumes a very good homogeneity of the electronic density of the 2DEG and so a very high quality of AlGaAs/GaAs heterostructures.

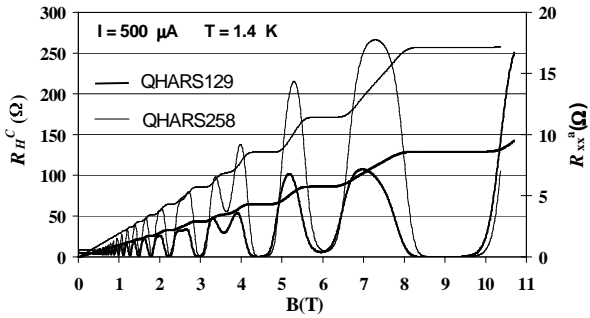

Fig. 6: $R_H$ and $R_{xx}^{a}$ in function of the magnetic flux density $B$ for QHARS129 and QHARS258 samples.

Fig. 6 shows, for both QHARS129 and QHARS258 samples, the whole evolution of the equivalent Hall resistance $R_H^{C}(B+)$ and the longitudinal resistance $R_{xx}^{a}(B+)$ as a function of the magnetic flux density $B$ at 1.4 K with a 500 µA supplying current. Very well defined plateaus of the Hall resistance and Shubnikov-de Haas oscillations are displayed. The carrier density and mobility of the 2DEG deduced from the analysis of $R_H^{C}$ and $R_{xx}^{a}$ at low magnetic field are in good agreement with heterostructure values: $n_s$=4.25×10$^{15}$ m$^{-2}$ and µ=50.8 T$^{-1}$. The resistance of current contacts is near 1 Ω, while that of voltage contacts is less than 0.5 Ω (the measurement resolution).

Within the $i$=2 plateau, the $R_{xx}^{a}$ dependence on $B$ of QHARS129 and QHARS258 samples supplied by a 100 µA current are respectively characterized by minimum flat regions of 0.4 T and 0.6 T width which progressively shrink with increasing current (see Fig. 7). However the Hall resistance plateaus remain perfectly flat for current values up to 2 mA. Just around the center of the flat region, the $V_{xx}^{a}$ voltage values are not symmetric with current reversal. Moreover, depending on the cooling down process it happens that the sign of the voltage sometimes changes around a defined value of $B$. The definition of $R_{xx}^{a}$ is thus delicate around its minimal value. Nevertheless, there exists a given magnetic flux density (~8.7 T) around which lowest absolute values are measured. Depending on the thermal cycle undergone by the samples, the minimal $R_{xx}^{a}$ values measured at the center of the plateau varies from 20 µΩ to 150 µΩ. These values should correspond, in the case of the QHARS129, to a mean longitudinal resistance value $\overline{r_{xx}^{a}} = 100 R_{xx}^{a}$ for a single Hall bar varying from 2 mΩ to 15 mΩ. These observations, and particularly the high values of $\overline{r_{xx}^{a}}$ compared to what is expected for a single Hall bar, put forward the contribution of the contacts, some of them being not perfect. As explained in the previous section, parasitic currents due to contact effects circulate in the voltage probes used to measure $R_{xx}^{a}$. This quantity consequently deviates with a second order term $O(\varepsilon^2)$ from the intrinsic longitudinal resistance of the sample $R_{xx}$ (equation 4).

$R_{xx}^{b}$ (equation 5) is the quantity measured in order to emphasize these contact effects on the longitudinal resistance. For QHARS129 sample, $R_{xx}^{b}$ values lower than 1 µΩ were found. They correspond to values of $\overline{r_{xx}^{b}} = 100 R_{xx}^{b}$ lower than 100 µΩ. This result, very different from the one previously found for $\overline{r_{xx}^{a}}$, is in perfect agreement with the equation (6) which indicates that $R_{xx}^{b}$ deviates from $R_{xx}$ by a third order term $O(\varepsilon^3)$ only. At the same time, it confirms the efficiency of the multiple connections to eliminate the contact resistance effects. Comparisons of $R_{xx}^{a}$ with $R_{xx}^{b}$ and of $R_H^{C}$ with $R_H^{D}$ for both directions of $B$ thus allow us to estimate the magnitude of $O(\varepsilon^2)$ and $O(\varepsilon^3)$ terms. They amount to some parts in 10$^6$ and some parts in 10$^9$ respectively. These results are consistent with a typical $\varepsilon r_H/2$ mean value of 10 Ω. The study of one 100 µm large Hall bar has indeed shown that some contact could have such high resistance values.

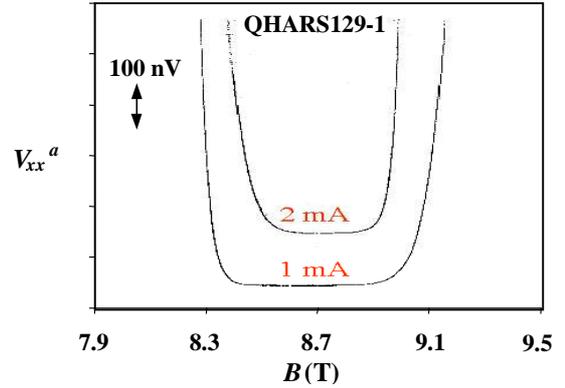

Fig. 7: $V_{xx}^{a}$ versus $B$ around $i$=2 plateau for two currents for QHARS129.

In order to validate the arrays for a metrological use, a fine characterization of the Hall resistance plateaus is needed. Accurate measurements of $R_H^{C}(B+)$ have been carried out for QHARS129 samples at the region where $R_{xx}^{a}$ presents a minimum corresponding to the $i$=2 plateau. Fig. 8 presents the fine shape of this plateau measured with a 400 µA current. For a first QHARS129-1 sample, $R_H^{C}(B+)$ values does not deviate from $R_K/200$ value by more than 5 parts in 10$^9$ (within the measurement uncertainty) over a range of about 0.6 T. That denotes a very good quantization, which allows a comfortable use of the standard in metrology. The same result is obtained on a second sample QHARS129-2 with a deviation lower than 7 parts in 10$^9$ over 0.4 T. Beyond these ranges, more important deviations are observed in agreement with the general shape reported in the inset of Fig. 8. Actually, near the magnetic field regions corresponding to a strong dissipation and especially on the high magnetic field side, a dip of resistance appears for $B(+)$. Superposition of $B(+)$ and $B(-)$ curves indicates that the sign of this deviation changes with the magnetic field reversal. This effect can be explained by a mixing of $R_H$ and $R_{xx}$ components in the strong dissipation region due to the 2DEG in-homogeneity[36,37]. According to this approach, the 2DEG in-homogeneity implies the existence of a transverse current component between opposite Hall probes which yields a linear coupling between the Hall and the longitudinal resistance. The sign and the value of the coupling coefficient then depend on the cool down process. The magnetic shift between the $R_{xx}^a$ minimum and the $R_H$ plateau center reinforced this hypothesis. Since $R_{xx}^a$ and $R_H$ quantities are measured at different probes, this magnetic shift may indeed be understood as a direct consequence of the existence of a notable 2DEG inhomogeneity. Another mechanism leads to a linear coupling between $R_{xx}$ and $R_H$: an effective probe misalignment results from the flowing of the voltage measurement current between diagonally opposite edges of the Hall probes. In this case, the coupling doesn't change sign by magnetic field reversal. In a general way, both mechanisms are needed to account for observations[37,38]. The physics of this coupling near the plateau center is in fact very complex because it has been observed that the mixing is highly dependent on the filling factor and current intensity[36,39].

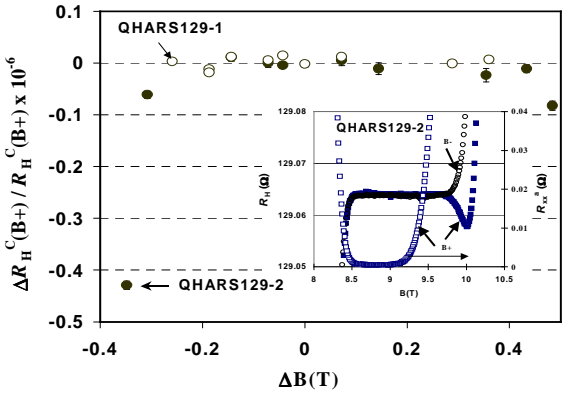

Fig. 8: $i=2$ $R_H$ plateau of two QHARS129 samples. Inset: enlarged view of the plateau for $B+$ and $B-$.

So, the behaviors of $R_H$ and $R_{xx}$ at the center of the $i=2$ plateau with current and temperature were studied (Fig. 9). For the QHARS129-1 sample, neither $R_H$ nor $R_{xx}^b$ vary noticeably when the current varies from 100 µA to 400 µA. $R_{xx}^b$ is equal to zero within the measurement uncertainty (~0.3 µΩ). $R_{xx}^b$ values lower than 1 µΩ correspond to intrinsic longitudinal resistance values $\overline{r_{xx}^b} \approx \overline{r_{xx}}$ lower than 100 µΩ. These low $\overline{r_{xx}}$ values indicate that the carrier transport is dissipationless like in high quality single Hall bars. For the QHARS129-2 sample, higher values of $R_{xx}^b$ were found: 20 µΩ at 1.3 K and with a supplying current of 400 µA. Nevertheless any variation of $R_H$ with the current can be observed as well. Measurements carried out on the QHARS129-2 sample show that $R_{xx}^b$ is approximately multiplied by a factor of 200 when the temperature T varies from 1.3 K to 3.5 K (inset: Fig. 9). This large increase must be related to the narrowness of the $R_{xx}^a$ minimum region, notably for the QHARS129-2 sample, compared to that of a single Hall bar. The 2DEG inhomogeneity randomly shifts the $r_{xx}$ minimum of the Hall bars composing the standard, and thus leads to an effective reduced dissipationless region. $R_{xx}^a$ also highly increases with temperature. In fact, the resistance values of $R_{xx}^a$ and $R_{xx}^b$ perfectly agree at 2.9 K and 3.5 K. As observed before, that's not the case at 1.3 K. This equality at high temperature is due to the relative decreasing contribution of $O(\varepsilon^2)$ to $R_{xx}^a$ when increasing temperature. The temperature range of measurement was not large enough to discriminate between activated or variable range hopping conduction[39-41]. Despite this exponential variation of $R_{xx}^b$ and similarly of $R_{xx}$, no systematic deviation of $R_H$ with temperature can be observed. Under these conditions one can deduce that the effective coupling between $R_H$ and $R_{xx}$ is very weak at the center of the plateau. Although the 2DEG inhomogeneity is likely to be responsible for enhanced $R_{xx}$ values, the whole QHARS might nevertheless benefit from a self-averaging of the coupling coefficients whose values would be randomly distributed from one Hall bar to the other.

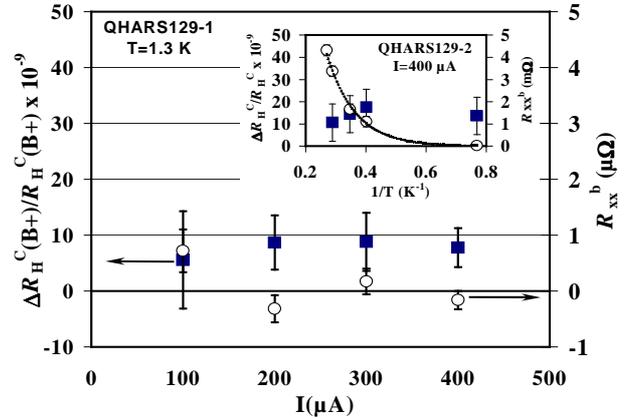

Fig. 9: $\Delta R_H/R_H(B+)$ and $R_{xx}^b$ versus current at 1.3 K for QHARS129-1 sample. Inset: $\Delta R_H/R_H(B+)$ and $R_{xx}^b$ versus 1/T for QHARS129-2 sample.

Finally, we have measured the quantization accuracy. Fig. 10 presents the results of the comparisons made with QHARS129-1 over a period of five months. Currents of 1 mA and 2 mA, corresponding respectively to 10 µA and 20 µA in a single Hall bar, were used for the direct comparisons. The indirect comparisons have been realized with a limited current value of 400 µA imposed by the maximum 50 mA current value supplied in the 1 Ω resistor. In each case, the relative deviation between the Hall resistance $R_H^C(B+)$ and its nominal value $R_K/200$ is smaller than $5\times10^{-9}$. The mean deviation is $0.1\times10^{-9}$ for the direct comparisons and $2.4\times10^{-9}$ for the indirect comparisons with combined standard relative uncertainties of the mean equal to $2\times10^{-9}$ and $3\times10^{-9}$ respectively. For the QHARS129-2 sample (see Table II), the relative discrepancy of $R_H^C(B+)$ directly measured in terms of $R_K/2$ is also low ($0.1\times10^{-9}$) but the combined standard relative uncertainty is larger ($10\times10^{-9}$). A notable negative relative deviation ($-9.6\times10^{-9}$) can be deduced from the indirect measurements. In that case, we think that the short time instability of the material standards during the measurements that spread over one day or more has contributed to the observed discrepancy. Indeed, the resistance evolution of our 1 Ω resistor is characterized by short time instability periods with peak to peak relative deviations reaching 2 parts in $10^8$. For the QHARS258 sample with a 500 µA current (10 µA in each Hall bar), only direct comparisons with $R_K/2$ have been carried out. The indirect comparisons based on the use of a 2 Ω transfer resistor have not yet been possible because the material standard available was not stable enough.

A peak to peak dispersion of about 2 parts in $10^8$ is observed on measurements. The mean deviation measured is slightly negative and equal to $-13 \times 10^{-9}$ with a combined standard relative uncertainty of $3.5 \times 10^{-9}$.

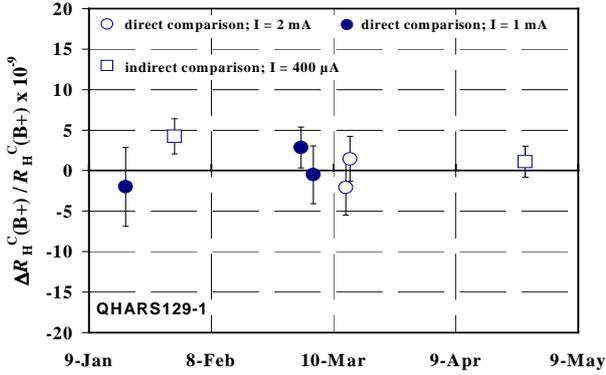

Fig. 10: $\Delta R_H/R_H(B+)$ versus time for QHARS129-1 sample.

## V. Discussion

Table II sums up mean deviations and respective uncertainties measured with QHARS. The results show that discrepancies of the Hall resistance are lower than 15 parts in $10^9$ in any cases.

| $\Delta R_H^C(B+)/R_H^C(B+)$ (combined standard uncertainty) $\times 10^{-9}$ | | |
|---|---|---|
| Samples | Direct | Indirect |
| QHARS129-1 | 0.1 (2) | 2.4 (3) |
| QHARS129-2 | 0.1 (10) | -9.6 (3) |
| QHARS258-2 | -13 (3.5) | |

Table II: sum up of measured deviations $\Delta R_H^C(B+)/R_H^C(B+)$.

For the QHARS129-1 sample, results are better. And the small deviations of $R_H$ obtained are entirely consistent with the very low values of $R_{xx}$ measured. For the QHARS129-2 and QHARS258-1 samples, a systematic negative deviation seems to appear depending on the kind of measurement (direct or indirect). But, only measurements with lower uncertainties could display such small deviations. All these quantum resistance standards are nevertheless characterized by a strong robustness with regard to magnetic field, temperature and current variations, that makes them in principle very useful for metrological applications. As single Hall bars, QHARS samples once cooled down are more or less good. It is consequently important to verify some properties like the existence of a flat $R_{xx}^a$ minimum, measure the $R_{xx}^b$ and the contact resistance values.

Let us now discuss some mechanisms able to alter the quality of QHARS and consequently imply some Hall resistance deviations, $R_{xx}$ increases, or parasitic effects. With regard to the sample QHARS129-2, higher $R_{xx}$ values have been measured. However, as demonstrated before, the coupling between $R_{xx}$ and $R_H$ seems too weak to induce notable discrepancy. Leakage currents in the sample itself might be responsible for Hall resistance deviations. But at low temperature, tunnel and activated leakage currents are assumed to be negligible. Test samples will be designed in the next fabrication project to test the electrical insulation and definitively reject the leakage hypothesis. Residual contact resistance effects may also be the cause of Hall resistance discrepancies. From measurements of $R_H^C$ and $R_H^D$ in positive and negative magnetic field, the third order $O(\varepsilon^3)$ contribution has been estimated to be up to 8 parts in $10^9$. As observed, the amplitude and the sign of the induced deviations are sample and cool down dependent. Contact effects are therefore likely to contribute above all to the time dispersion of the measurements. As explained before, a part of the measurement scattering may be also due to the short time instability of the 1 $\Omega$ material resistor.

We must finally discuss the influence of the connections on the 2DEG. The gold links between voltage probes which cross the Hall bar channels (see figure 2c) may indeed act as electronic density modulation gates. Gates have been widely used to create local barriers across Hall bars in order to test the edge-state theory of the transport[42]. By varying the carrier density, the filling factor $\nu$ ($n_s h/eB$) in the gate region is changed and thus becomes different from that in the undisturbed part. The number of edge-states being related to the filling factor, the gate therefore allows modulation of the reflection of the edge-states. Depending on the sign of the carrier concentration variation, the longitudinal resistance along a Hall bar crossed by a barrier can present quantized or non-quantized plateau values versus gate voltage. But these results only apply to systems supplied with low currents and characterized by gate voltages much higher than Hall voltages. When the current is increased the plateaus are altered, non linear effects related to the potential distribution in the gate region appear. To quantify this density modulation effect, we must firstly determine the capacitance C per surface unit between the 2DEG and the gold connections. The capacitance per surface unit between two infinite metallic plates separated by an homogeneous dielectric material of thickness d and relative dielectric constant $\varepsilon_r$ is given by $C = \varepsilon_0 \varepsilon_r/d$. Assuming the following relative dielectric constants for the layers ($\varepsilon_{GaAs} = 13$, $\varepsilon_{SiO2} = 4$ and $\varepsilon_{Si3N4} = 7$), the capacitance $C_A$ and $C_B$ between the 2DEG and respectively the connections A and B are given by $C_A \sim \varepsilon_0 /( d_2/\varepsilon_{Si3N4} + d_1/\varepsilon_{GaAs}) \sim 3.3 \times 10^{-4}$ Fm$^{-2}$ and $C_B \sim \varepsilon_0 /( d_3/\varepsilon_{SiO2} + d_2/\varepsilon_{Si3N4} + d_1/\varepsilon_{GaAs}) \sim 8.7 \times 10^{-5}$ Fm$^{-2}$. Assuming this model, the associated variations of the local electronic density $n_s$ induced by the capacitance effect under the connections are given by $\delta n_{sA}(x)/\Delta V(x) \sim C_A/e \sim 2.1 \times 10^{15}$ m$^{-2}$V$^{-1}$ and $\delta n_{sB}(x)/\Delta V(x) \sim C_B/e \sim 5.4 \times 10^{14}$ m$^{-2}$V$^{-1}$, with $\Delta V(x)$ the voltage difference between the 2DEG and the connection at some position x. For $\Delta V(x)=0.13$ V, which corresponds to the Hall voltage of a QHARS129 sample supplied with a 1 mA current, one obtains $\delta n_{sA}(x) \sim 2.7 \times 10^{14}$ m$^{-2}$ and $\delta n_{sB}(x) \sim 7 \times 10^{13}$ m$^{-2}$. For our samples, these values correspond to relative variations of the carrier density equal to 6.4 % and 1.6 % respectively. The induced variations of $\nu$ being far from reaching integer values, a low reflection of edge states is thus expected in our case. Since for 1 mA current the electrochemical potential is more than 8 times the cyclotron energy ($\hbar \omega_c \sim 16$ meV), the edge-state theory is not valid, bulk currents exist and the gate effects must be consequently altered. Finally, by construction the connections are at a potential equal to that of the edge at which they are connected. In the quantized regime the electric field is no longer homogeneous and the potential mainly drops near the sample edges even for so large currents. For one particular link crossing the sample and connected at a given edge, electrons near this edge are approximately at the same potential as the link while electrons near the opposite edge are at a potential different from about the Hall voltage (see figure 2b). The filling factor modulation therefore is not homogeneous across the channel since it doesn't exist at one edge. The reflection of an edge state by the gate effect should be therefore drastically reduced. As detailed before very low longitudinal resistance values have been measured, notably for the QHARS129-1 sample, confirming that the gate effects are limited. However, not only higher values have been obtained for the second sample, but also the maximum current with which samples can be supplied stays limited at a quite lower value (2 mA) than expected (4 mA). Gate effects whose amplitude could be modulated by the natural and sample dependent density inhomogeneity can survive and affect the quality of our samples. Even if the connections don't intercept the Hall voltage line, spatial filling factor variations can modify the current distribution from that would be expected in a perfect Hall bar geometry and which is essential to obtain the best accurate quantization.

These observations allow us to deduce some technical guidelines for improving the QHARS. Firstly, the samples must be preferably designed in such a way that connections don't cross the Hall bar channels, or cross them with reduced charging effects on the 2DEG. This is likely to cancel potential systematic deviations from the nominal quantized Hall resistance. An increase of the maximum supply current may also be expected. Samples based on Hall bars with larger channels will work with higher currents. For samples characterized by these carrier mobility and density parameters, critical currents indeed seem to increase linearly with the channel width[43-45]. However, the increase of the surface of the QHARS due to the channel enlargement goes against the global homogeneity of the electronic density and of the magnetic field. A compromise must be reached. Samples with quadruple connections must also be tested. Moreover, in case of parallel connections every voltage probe must be connected in order that the sample be unchanged when reversing the magnetic field. Measurements with positive and negative magnetic field might allow us to cancel some deviations. The samples described in this paper are based on the use of two insulating layers to separate the two levels of voltage connections, and this design implies the need of an increasing number of insulating layers to add new connections. The crossing of voltages probes can be avoided by the use of lateral electrical relays placed on each side of the chip. In that case, only one insulating layer is necessary and consequently the fabrication process is simplified. A way to limit the chip size consists in reversing the sign of the current injected from one Hall bar to the following one in the vertical direction. Indeed, this allows us to connect directly the bottom side probes of one Hall bar to the top side probes of the following Hall bar. Although probes must be connected diagonally, a lot of the space is saved. This method has another advantage: the geometry of the connections is similar for the connections in series and in parallel. For instance, an array with a nominal resistance value near 100 Ω can be made in the following way: 129 Hall bars and a system made of 16 Hall bars placed in series are connected in parallel. One obtains a standard whose nominal resistance is equal to $16R_K/4130 \sim 100(1+11.8935\times 10^{-6})$. Finally, a device equipped with a QHARS and a classical Hall bar should allow a simplified way to realize direct comparisons of quantum resistors.

## VI. Conclusion

We have developed new QHE standards with resistance values as low as $R_K/200$. Our study demonstrates the efficiency of the multiple connections technique to construct artificial macroscopic QHE standards that behave like single Hall bars. For specific good samples, we have indeed shown that the relative accuracy of the quantized resistance is within 5 parts in $10^9$. Moreover, QHARS are robust with regard to current and temperature variations. These properties might be based on a Hall bars self-averaging principle. QHARS are thus likely to open new metrological applications: they can be used as standards in combination with conventional bridges or as handy transportable resistors for international comparisons. From this study we have deduced some technical guidelines for designing improved QHARS notably able to work with higher currents. Such a new generation of QHARS has already been fabricated and is presently under test.

## VII. Acknowledgment

The authors are particularly grateful to G. Trapon for fruitful discussions and critical reading of the manuscript.

## VIII. Appendix

The notations have been defined in section III. They are also indicated on Fig. 5. We define the mean values $\overline{G}$ and $\overline{GH}$ by :

$$\overline{G} = \frac{1}{N}\sum_{q=1}^{N} G_q , \quad \overline{GH} = \frac{1}{N}\sum_{q=1}^{N} G_q H_q$$

The conservation of the currents in the circuit may be expressed by:

$$\overline{i} = \frac{I}{N}, \quad \overline{i'} = 0, \quad \overline{i''} = 0, \quad \overline{j} = \frac{I}{N}, \quad \overline{j'} = 0, \quad \overline{j''} = 0 \quad (7)$$

Let us start with the case of a positive magnetic field. Applying the Kirchhoff's laws to the closed loop (ABCDA) defined between the Hall bars numbered 1 and q, one obtains :

$$-i_1\varepsilon_{11a} + i_q\varepsilon_{q1a} - i'_1(2+\varepsilon_{11b}) + i'_q(2+\varepsilon_{q1b}) = 0 \quad (8)$$

The average of equations over q values varying between 1 and N gives:

$$\overline{i\varepsilon_{1a}} - i_1\varepsilon_{11a} + \overline{i'(2+\varepsilon_{1b})} - i'_1(2+\varepsilon_{11b}) = 0 \quad (9)$$

This method of calculation can be generalized to any q value (and not only 1) and applied to six closed loops. The additional use of equations (7) gives the following set of equations: $\forall q$

(loop ABCD):
$$\overline{i\varepsilon_{1a}} - i_q\varepsilon_{q1a} + \overline{i'\varepsilon_{1b}} - i'_q(2+\varepsilon_{q1b}) = 0 \quad (10)$$

(loop CDEFC):
$$\left[\overline{i\beta} - i_q\beta_q\right] - \left[\overline{i'(\beta+\varepsilon_{1b})} - i'_q(\beta_q+\varepsilon_{q1b})\right] + \left[\overline{i'\varepsilon_{2b}} - i'_q(2+\varepsilon_{q2b})\right] = 0 \quad (11)$$

(loop EFF'E'E left way):
$$2\left[\frac{I}{N} - i_q\right] + 2\left[i'_q\right] - \left[\overline{i'\varepsilon_{2b}} - i'_q(2+\varepsilon_{q2b})\right] + \left[\overline{j'\varepsilon_{2d}} - j'_q\varepsilon_{q2d}\right] = 0 \quad (12)$$

(loop EFF'E'E right way):
$$2\left[\frac{I}{N}-j_q\right]-2\left[j'_q\right]-\left[\overline{i''\varepsilon_{2b}}-i''_q\varepsilon_{q2b}\right]+\left[\overline{j''\varepsilon_{2d}}-j''_q(2+\varepsilon_{q2d})\right]=0 \quad (13)$$

(loop C'D'E'F'C'):
$$2\left[\frac{I}{N}-j_q\right]-2\left[j'_q\right]-\left[\overline{i''\varepsilon_{2b}}-i''_q\varepsilon_{q2b}\right]+\left[\overline{j''\varepsilon_{2d}}-j''_q(2+\varepsilon_{q2d})\right]=0 \quad (14)$$

(loop A'B'C'D'A'):
$$\overline{j\varepsilon_{3c}}-j_q\varepsilon_{q3c}-\left[\overline{j'\varepsilon_{3d}}-j'_q(2+\varepsilon_{q3d})\right]=0 \quad (15)$$

The set of equations (13), (14), (15) is similar to the set (10), (11), (12). The currents $j_q$, $j'_q$, $j''_q$ are consequently determined in the same way as currents $i_q$, $i'_q$, $i''_q$. Let us express $i_q$ by:

$$i_q=\frac{I}{N}\left[1+O_q(\varepsilon,\beta)\right] \quad (16)$$

$\frac{I}{N}$ being the $i_q$ value when the contact or longitudinal resistance values are null or equal between them. $O_q(\varepsilon,\beta)$ is a generic function in orders of $\varepsilon$ and $\beta$ equal or higher than 1. Introducing (16) in equation (10), we obtain the following expression for $i'_q$:

$$i'_q=\frac{I}{N}\left\{\frac{\left[\overline{\varepsilon_{1a}}-\varepsilon_{q1a}\right]}{2}+O_q(\varepsilon^2,\varepsilon\beta,\beta^2)\right\} \quad (17)$$

Using (17) in equation (11), the following result comes:

$$i''_q=\frac{I}{N}\left\{\frac{\left[\overline{\beta}-\beta_q\right]}{2}+O_q(\varepsilon^2,\varepsilon\beta,\beta^2)\right\} \quad (18)$$

$i''_q$ contains no $\varepsilon$ first order term. A similar result is also obtained for $j''_q$.

Using (17) and (18) in equation (12) and the fact that $j''_q$ contains no zero order term, we deduce the first order term of $i_q$:

$$i_q=\frac{I}{N}\left\{1+\frac{1}{2}\left[\overline{\varepsilon_{1a}}-\varepsilon_{q1a}+\overline{\beta}-\beta_q\right]+O_q(\varepsilon^2,\varepsilon\beta,\beta^2)\right\} \quad (19)$$

Coming back to the equation (11) and introducing results (17) and (19), we determine the expression of $i''_q$ at the second order.

$$i''_q=\frac{I}{N}\left\{\begin{array}{l}\frac{\left[\overline{\beta}-\beta_q\right]}{2}+\frac{1}{4}\left[\left(\overline{\beta^2}-\overline{\beta}^2\right)-\beta_q\left(\overline{\beta}-\beta_q\right)\right]\\+\frac{1}{4}\left[\left(\overline{\beta\varepsilon_{2b}}-\overline{\beta}\overline{\varepsilon_{2b}}\right)-\varepsilon_{q2b}\left(\overline{\beta}-\beta_q\right)\right]\\+\frac{1}{4}\left[\varepsilon_{q1b}\left(\overline{\varepsilon_{1a}}-\varepsilon_{q1a}\right)-\left(\overline{\varepsilon_{1a}\varepsilon_{1b}}-\overline{\varepsilon_{1a}}\overline{\varepsilon_{1b}}\right)\right]+O_q(\varepsilon^3,\varepsilon\beta,\beta\varepsilon^2,\beta^3)\end{array}\right\} \quad (20)$$

Using the same method of calculation, similar results are obtained for $j_q$, $j'_q$, $j''_q$.

$$j_q=\frac{I}{N}\left\{1+\frac{1}{2}\left[\overline{\varepsilon_{3c}}-\varepsilon_{q3c}+\overline{\beta}-\beta_q\right]+O_q(\varepsilon^2,\varepsilon\beta,\beta^2)\right\} \quad (21)$$

$$j'_q=\frac{-I}{N}\left\{\frac{\left[\overline{\varepsilon_{3c}}-\varepsilon_{q3c}\right]}{2}+O_q(\varepsilon^2,\varepsilon\beta,\beta^2)\right\} \quad (22)$$

$$j''_q=\frac{-I}{N}\left\{\begin{array}{l}\frac{\left[\overline{\beta}-\beta_q\right]}{2}+\frac{1}{4}\left[\left(\overline{\beta^2}-\overline{\beta}^2\right)-\beta_q\left(\overline{\beta}-\beta_q\right)\right]\\+\frac{1}{4}\left[\left(\overline{\beta\varepsilon_{2d}}-\overline{\beta}\overline{\varepsilon_{2d}}\right)-\varepsilon_{q2d}\left(\overline{\beta}-\beta_q\right)\right]\\+\frac{1}{4}\left[\varepsilon_{q3d}\left(\overline{\varepsilon_{3c}}-\varepsilon_{q3c}\right)-\left(\overline{\varepsilon_{3c}\varepsilon_{3d}}-\overline{\varepsilon_{3c}}\overline{\varepsilon_{3d}}\right)\right]+O_q(\varepsilon^3,\varepsilon\beta,\beta\varepsilon^2,\beta^3)\end{array}\right\} \quad (23)$$

The knowledge of currents allows to deduce the resistance expressions in positive magnetic field:

$$R_H^C(B+)=\frac{r_H}{N}\left\{1+\frac{1}{2}\left[\left(\overline{\frac{j''}{j}}\right)\varepsilon_{2d}-\left(\overline{\frac{i''}{i}}\right)\varepsilon_{2b}\right]\right\} \quad (24)$$
$$=\frac{r_H}{N}\left\{1+O(\varepsilon\beta,\varepsilon^3,\beta\varepsilon^2,\varepsilon\beta^2,\beta^3)\right\}$$

$$R_{xx}^a(B+)=\frac{r_H}{2N}\left\{\left[\left(\overline{\frac{i'}{i}}\right)\beta\right]-\left[\left(\overline{\frac{i''}{i}}\right)(\beta+\varepsilon_{1b})\right]+\left[\left(\overline{\frac{i''}{i}}\right)\varepsilon_{2b}\right]\right\} \quad (25)$$
$$=\frac{r_H}{2N}\left\{\overline{\beta}+O(\varepsilon^2,\beta\varepsilon,\beta^2)\right\}$$

$$R_H^D(B+)=\frac{r_H}{N}\left\{\begin{array}{l}1+\frac{1}{2}\left[\left(\overline{\frac{j''}{j}}\right)\varepsilon_{3d}-\left(\overline{\frac{i''}{i}}\right)\varepsilon_{1b}\right]\\+\frac{1}{2}\left[\left(\overline{\frac{i'}{i}}\right)\beta+\left(\overline{\frac{j'}{j}}\right)\beta-\left(\overline{\frac{i''}{i}}\right)\beta+\left(\overline{\frac{j''}{j}}\right)\beta\right]\end{array}\right\} \quad (26)$$
$$=\frac{r_H}{N}\left\{1+\overline{\beta}+O(\varepsilon^2,\beta\varepsilon,\beta^2)\right\}$$

The central Hall resistance $R_H^C(B+)$ doesn't contain second order terms, except $\beta\varepsilon$. The quantity $\frac{r_H}{2N}\overline{\beta}$ which appears in the expressions $R_{xx}^a(B+)$ and $R_H^D(B+)$ represents the first order term of the equivalent longitudinal resistance of the sample. The resolution of equations (10-12) in case of zero $\varepsilon$ values allows to obtain the exact quantity:

$$R_{xx}(B+)=\frac{r_H}{N}\left[\frac{\overline{\frac{\beta}{2+\beta}}}{1-\overline{\frac{\beta}{2+\beta}}}\right] \quad (27)$$

When contact effects are taken into account $\varepsilon^2$ add to these quantities. Results are different for $B-$ because the geometry of the connections is not symmetric by magnetic field reversal. So let us reverse the sample to determine the behavior for $B-$. Starting from one side of the sample, the first and the second potential connections respectively link central terminals and opposite terminals. Consequently, $\varepsilon$ terms add to the currents $i''_q$ and $j''_q$ while currents $i'_q$ and $j'_q$ contain $\varepsilon^2$ terms only.

## **References**

[1] K. von. Klitzing, G. Dorda and M. Pepper, Phys. Rev. Lett. **45**, 494 (1980).
[2] R. B. Laughlin, Phys. Rev. B **23**, 5632 (1981).
[3] B. I. Halperin, Phys. Rev. B **25**, 2185 (1982).

[4] Comité International des Poids et Mesures, « Représentation de l'ohm à partir de l'effet Hall quantique », Recommandation 2 (CI-1988), 77th session, 1988.

[5] Comité consultatif d'électricité et magnétisme (CCEM), Report of the 22nd Meeting, (2000).

[6] P. J. Mohr, B. N. Taylor, J. Phys. Chem. Ref. Data **28**, n°6 (1999).

[7] G. Trapon, O. Thénenot, J. C. Lacueille, W. Poirier, H. Fhima and G. Genevès, IEEE Trans. Instrum. Meas. **50**, 572 (2001).

[8] M. Cage et al, IEEE Trans. Instrum. Meas. **38** (2), 284 (1989).

[9] A. Jeffery, R. E. Elmquist, J. Q. Shields, L. H. Lee, M. E. Cage, S. H. Shields, and R. F. Dziuba, Metrologia **35** (2), 83 (1998).

[10] F. Delahaye et al, Compte rendu CCE, Appendix E4, E119 (1988).

[11] U. Klass, W. Dietsche, K. von Klitzing, and K. Ploog, Z. Phys. B **82**, 351 (1991).

[12] D. A. Syphers, F. F. Fang, P. J. Stiles, Surface Science **142**, 208 (1984).

[13] F. F. Fang, P. J. Stiles, Phys. Rev. B **29**, 3749 (1984).

[14] F. Delahaye, J. Appl. Phys. **73**, 7915 (1993).

[15] B. W. Ricketts and P. C. Kemeny, J. Phys. D **21**, 483 (1988).

[16] M. Büttiker, Phys. Rev. B **38**, 9375 (1988).

[17] P. F. Fontein, P. Hendricks, F. A. P. Blom, J. H. Wolter, and L. J. Gilling, Surf. Sci. **263**, 91 (1992).

[18] R. Knott, W. Dietsche, K. von, Klitzing, K. Eberl, and K. Ploog, Semincond. Sci. Techno. **10**, 117 (1995).

[19] W. Dietsche, K. von Klitzing, and K. Ploog, Surf. Sci. **361**, 289 (1996).

[20] E. Yahel, A. Tsukernik, A. Palevski, and H.Shtrikman, Phys. Rev. Lett. **81**, 5201 (1998).

[21] F. Delahaye, T. J. Witt, F. Piquemal, G. Genevès, IEEE Trans. Instrum. Meas. **44**, 258 (1995).

[22] F. Piquemal, J. Blanchet, G. Genevès, J. P. André, IEEE Trans. Instrum. Meas. **48**, 296 (1999).

[23] Mutual Recognition Arrangement (MRA), 21st CGPM, http://www.bipm.org (1999)

[24] A. Satrapinsky, H. Seppä, B. Schumacher, P. Warneke, F. Delahaye, W. Poirier, and F. Piquemal, IEEE Trans. Instrum. Meas. **50**, 238 (2001).

[25] A. D. Inglis, IEEE Trans. Intrum. Meas. **48**, 289 (1999)

[26] F. Delahaye, D. Reymann, IEEE Trans. Meas. **IM34**, 316 (1985).

[27] F. Gay, F. Piquemal, G. Genevès, Rev. Sci. Instrum. **71**, 4592 (2000).

[28] K. Pierz, B. Schumacher, IEEE Trans. Intrum. Meas. **48**, 293 (1999)

[29] K. Harvey, Rev. Sci. Instrum. **43**, 1626 (1972).

[30] F. Piquemal, B. Etienne, J. P. André, J. N. Patillon, IEEE Trans. Meas. **40**, 234 (1991).

[31] F. Piquemal, bulletin du Bureau National de Métrologie **116**, (1999).

[32] W. Poirier, F. Piquemal, H. Fhima, N. Bensaid, G. Genevès, J. P. André, in Proc. 9st Congr. Int. Métrologie-Bordeaux, 138 (1999).

[33] S. Komiyama, H. Hirai, S. Sasa, and T. Fujii, Surf. Sci. **229**, 224 (1990).

[34] B. Jeckelmann, B. Jeanneret, D. Inglis, Phys. Rev. B **55**, 13124 (1997).

[35] B. Jeckelmann, B. Jeanneret, IEEE Trans. Instrum. Meas. **46**, 276 (1997).

[36] D. Dominguez, PhD thesis CNAM (1987).

[37] W. van der Wel, PhD thesis Delft University (1987).

[38] M. E. Cage, Phys. Rev. B **30**, 2286 (1984).

[39] M. Furlan, Phys. Rev. B **57**, 14818 (1998).

[40] D. G. Polyakov, B. I. Shklovskii, Phys. Rev. B **48**, 11167(1993).

[41] S. Koch, R. J. Haug, K. von Klitzing, K. Ploog, Semicond. Sci. Technol. **10**, 209 (1995).

[42] R. J. Haug, Semicond. Sci. Technol. **8**, 131 (1993).

[43] A. Boisen, P. B∅ggild, A. Kristensen, and P. E. Lindelof, Phys. Rev. B **50**, 1957 (1994).

[44] N. Q. Balaban, U. Meirav, and H. Shtrikman, Phys. Rev. B **52**, R5503 (1995).

[45] B. Jeckelmann, A. Rüfenacht, B. Jeanneret, F. Overney, A. von. Campenhausen, and G. Hein, IEEE Trans. Intrum. Meas. **50**, 218 (2001).